\documentclass{article}
\usepackage{spconf,amsmath,graphicx}
\usepackage{multirow}
\usepackage{booktabs}
\usepackage{tabularx}

\title{MACCIF-TDNN: MULTI ASPECT AGGREGATION OF CHANNEL AND CONTEXT INTERDEPENDENCE FEATURES IN TDNN-based SPEAKER VERIFICATION}
\name{$Fangyuan$ $Wan{g^1}$, $Zhigang$ $Son{g^2}$, $Hongchen$ $Jian{g^1}$, $Bo$ $X{u^1}$}
\address{${}^{1}Institute$ $of$ $automation$, $Chinese$ $Academy$ $of$ $Sciences$ $,Beijing$ $,China$\\
${}^{2}Faculty$ $of$ $Information$ $Technology,$ $Beijing$ $University$ $of$ $Technology$ $,Beijing$ $,China$\\
}
\begin{document}
\bibliographystyle{ieeetr}
%
\maketitle
\begin{abstract}
Most of the recent state-of-the-art results for speaker verification are achieved by X-vector and its subsequent variants. In this paper, we propose a new network architecture which aggregates the channel and context interdependence features from multi aspect based on Time Delay Neural Network (TDNN). Firstly, we use the SE-Res2Blocks as in ECAPA-TDNN to explicitly model the channel interdependence to realize adaptive calibration of channel features, and process local context features in a multi-scale way at a more granular level compared with conventional TDNN-based methods. Secondly, we explore to use the encoder structure of Transformer to model the global context interdependence features at an utterance level which can capture better long term temporal characteristics. Before the pooling layer, we aggregate the outputs of SE-Res2Blocks and Transformer encoder to leverage the complementary channel and context interdependence features learned by themself respectively. Finally, instead of performing a single attentive statistics pooling, we also find it beneficial to extend the pooling method in a multi-head way which can discriminate features from multiple aspect. The proposed MACCIF-TDNN architecture can outperform most of the state-of-the-art TDNN-based systems on VoxCeleb1 test sets.
\end{abstract}
\begin{keywords}
speaker verification, speaker recognition, x-vector , transformer encoder, ECAPA
\end{keywords}
\section{Introduction}
\label{sec:intro}

Text independent speaker verification verifies the identity of a speaker by judging whether a speech segment is from the speaker. The flexibility of the technology is not limited by this text content, so it has extensive application scenarios. Most state-of-the-art speaker verification systems are based on Time delay neural network (TDNN), therefore improving TDNN-based network structure has always been an active area of research. Enhancing the temporal context modeling capability can be very effective which has been proved by introducing LSTM or GRU layer before the pooling layer in the typical TDNN \cite{1,2}. \cite{3} uses the Transformer \cite{4} encoder module to replace the TDNN network module in the conventional X-vector system \cite {5, 6} which can capture better global temporal characteristics. Besides,puts more emphasis on channel interdependence has also shown impressive performance improvement in VoxCeleb test sets and the 2019 VoxCeleb Speaker Recognition Challenge\cite {7,8}.

In this work we propose the MACCIF-TDNN to further enhance the TDNN architecture and statistics pooling layer. we use the SE-Res2Blocks as in ECAPA-TDNN \cite {7} to explicitly model the channel interdependence to realize adaptive calibration of channel features, and process context features in a multi-scale way at a more granular level compared with conventional TDNN-based methods. And, we explore to use the encoder structure of Transformer to model the global context interdependence features at an utterance level. Besides, we also find it beneficial to extend the pooling method in a multi-head way which can discriminate features from multiple aspect. We expect that this arrangement would aggregate channel interdependece, local and global context interdependence features from multi aspect. To the best of our knowledge, this is the first work that integrates SE-Res2Blocks, Transformer encoder and multi-head attentive statistics pooling in TDNN-based network.

\section{Baseline system artichitectures}
\label{sec:rw}

\subsection{ECAPA-TDNN}
\label{sec:ecapa}
The ECAPA-TDNN \cite{7} architecture is an enhanced version of the conventional X-vector system \cite {5,6}. It integrates a Res2Net module to enhance the central volume layer and constructs a hierarchical residual connection to handle multi-scale features. It also introduces 1-dimensional TDNN-specific SE-blocks \cite{9} which rescales the intermediate time context bound frame-level features per channel according to global utterance properties. The pooling layer uses a channel-and context-dependent attention mechanism to attend to different speaker characteristic properties at different time steps for each feature map. Finally, Multi-layer Feature Aggregation (MFA) provides additional complementary information for the statistics pooling by concatenating the final frame-level features with the intermediate features of previous layers. The architecture is trained with the AAM-softmax \cite{10,11} loss. Additionally, \cite{8} introduces a 2D convolutional stem in a strong ECAPA-TDNN baseline to transfer some of the strong characteristics of a ResNet based model to this hybrid CNN-TDNN architecture. 

\subsection{S-vector}
\label{sec:sv}
The S-vector \cite{3}proposes an self-attention based alternative to the X-vectors system \cite {5,6}. The TDNN of X-vector system, which has finite context, is replaced  with the encoder module of the Transformers. This arrangement would capture better speaker characteristics due to unrestricted context. Also, self-attention is built on the dot product between frames, so it can capture the similarities across an utterance efficiently. 
\subsection{R-vector}
\label{sec:rv}
The R-vector system proposed in \cite{12}. It is based on the ResNet18 and ResNet34 implementations of the successful ResNet architecture. The convolutional frame layers of this network process the features as a 2-dimensional signal before collecting the mean and standard deviation statistics in the pooling layer.

\section{Proposed system architectures}
\label{sec:proposed}

In this section, we examine some of the limitations of the TDNN-based methods and incorporate potential solutions into our MACCIF-TDNN architecture. The following subsections will focus on the complementary features aggregation and multi-head attentive statistic pooling. An overview of the complete architecture is given by Fig.1.
 
\begin{figure}[htb]
\begin{minipage}[b]{1.0\linewidth}
  \centering
  \centerline{\includegraphics[scale=1.1]{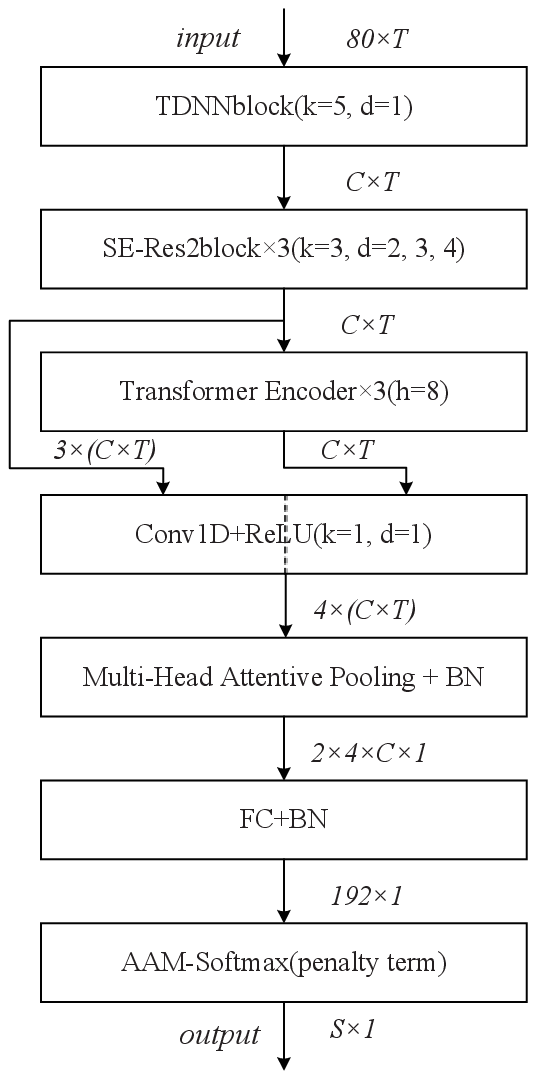}}
  \caption{Network topology of the MACCIF-TDNN. We denote $h$ for the number of heads, $k$ for kernel size and $d$ for dilation spacing of the Conv1D layers. $C$ and $T$ correspond to the channel and temporal dimension of the intermediate feature-maps respectively.$S$ is the number of training speakers. BN stands for Batch Normalization \cite{11} and the non-linearities are Rectified Linear Units (ReLU) unless specified otherwise.}
\end{minipage}
\end{figure}
 \subsection{Aggregation of complementary features}
\label{ssec:ucf}

The temporal context in the TDNN-based architecture is limited to several frames. As the network apparently benefits from longer term dependence\cite{1}, we argue it could be beneficial to strengthen the capability of global temporal context modeling. For this purpose we introduce the encoder structure of Transformer to our architecture. Different from the RNN networks, Transformer relies entirely on an attention mechanism to draw global dependence. Therefore, the advantage of the Transformer encoder structure is that it is not restricted to finite context and attends to all frames in every time step. However, encoder structure does not learn the characteristics of channel interdependence, and lacks the ability of multi-scale temporal context modeling. On the other hand, the SE-Res2Block integrates SE block and Res2Net, which not only models the channel interdependence, but also enhances the multi-scale local context model ability. Therefore, we introduce the SE-Res2Block with the Transformer encoder module.

We aggregate the complementary features extracted by these two parts as in Fig.2. For each frame we concatenate the output feature maps of the SE-Res2Blocks and the encoder layer. After the Multi-layer Feature Aggregation(MFA), a dense layer processes the concatenated information to generate the features for the attentive statistics pooling. To further aggregate multi-layer information we use the output of all preceding layer and initial convolutional layer as input for each frame layer which is implemented by defining the residual connection in each frame layer as the sum of the outputs of all the previous layers.

\begin{figure}[t]
\begin{minipage}[b]{1.0\linewidth}
  \centering
  \centerline{\includegraphics[scale=1.1]{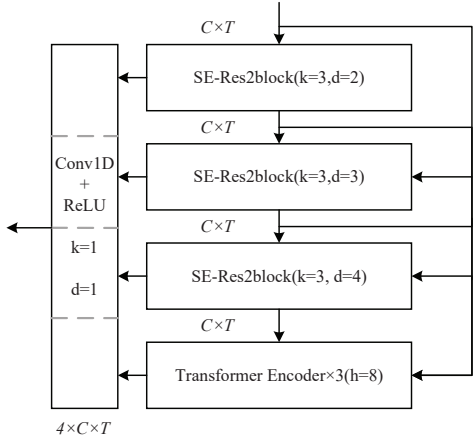}}
  \caption{The feature aggregation mechanism between the frame layers in MACCIF-TDNN. $C$ and $T$ correspond to the channel and temporal dimension of the intermediate feature-maps respectively.}
\end{minipage}
\end{figure}

\subsection{Multi-Head channel and context-dependent statistics pooling}
\label{ssec:subhead}
In recent TDNN-based architectures, the attention mechanism is integrated into the temporal pooling \cite{13,14}. \cite{7} extend this temporal attention mechanism even further to the channel dimension,which enables the network to focus more on speaker characteristics that do not activate on identical or similar time instances. Inspired by Vector-Based Attentive Pooling \cite{15}, we argue that it  might be beneficial to extend the channel and context-dependent statistics pooling in a multi-head way for better speaker embeddings from multiple aspect. 

The channel and context-dependent attention mechanism in \cite{7} defines as:
\begin{equation}
A = \tau {({W_2}f({W_1}{H^T} + {b_1}) + {b_2})^T}
\end{equation}
where weight over all frames can be represented as a matrix ${\rm{ }}A = \left[ {{a_1}, \cdots ,{a_T}} \right] \in {R^{T \times C}}$ with ${\rm{ }}{a_t} = \left[ {{a_{t1}}, \cdots ,{a_{tC}}} \right]$. The ${a_{tc}}$ is the self-attention score which represents the importance of each frame given the channel.

To discriminate speakers from multiple aspect, we extend the pooling method in a multi-head way as following equation:
\begin{equation}
{A^i} = \tau {({W_2}^if({W_1}^i{H^T} + {b_1}^i) + {b_2}^i)^T}
\end{equation}
where ${W_1} \in {R^{R \times C}}$ and ${W_2} \in {R^{C \times R}}$ are weight matrices; ${b_1} \in {R^{R}}$ and ${b_2} \in {R^{C}}$ are bias items; ${A^i}$ is the vectorial weight matrix generated by $i$-th attention head. The weighted mean vector ${\mu ^i}$ and weighted standard deviation vector ${\sigma ^i}$ generated by $i$-th attention head can be obtained by the following equations:
\begin{equation}
{\mu ^i} = \sum\limits_{t = 1}^T {\alpha _t^i}  \odot {h_t}
\end{equation}
\begin{equation}
{\sigma ^i} = \sqrt {\sum\limits_{t = 1}^T {\alpha _t^i}  \odot {h_t} \odot {h_t} - {\mu ^i} \odot {\mu ^i}}
\end{equation}

The final output of the pooling layer is given by concatenating the weighted mean vector ${\mu ^i}$ and weighted standard deviation vector ${\sigma ^i}$ in all I attention heads:

\begin{equation}
S = [{\mu ^1}; \cdots ;{\mu ^I};{\sigma ^1}; \cdots ;{\sigma ^I}]
\end{equation}

To ensure that each head can collect dissimilar information, a penalty term P is added to the loss function, the penalty term encourages the diversity of the attention matrices across different heads of attention: 
\begin{equation}
P = \rho \sum\limits_{i = 1}^I {\sum\limits_{j = i + 1}^I {\max (\lambda  - F{{({A^i} - {A^j})}^2},0)} }
\end{equation}
where $\lambda$ and $\rho$ are corresponding hyper-parameters; $f( \cdot )$ represents the Frobenius norm of matrix.

\section{Experiments}
\label{sec:ecapa}
\subsection{Dataset and Data Augmentation}
\label{sec:datab}
We train models with the development set of VoxCeleb1 and the development set of voxceleb-2 respectively. VoxCeleb1 \cite{16} development dataset contains over 100000 utterances for 1211 celebrities, extracted from videos uploaded to YouTube. VoxCeleb-2 \cite{17} development dataset contains over 1 million utterances for 5994 celebrities, extracted from videos uploaded to YouTube. In order to generate additional augmented copies, we augment the data with reverberation (The RIR dataset\cite{18}) and noise (MUSAN dataset\cite{19}). And the remaining three augmentations are generated with the open-source SoX (tempo up, tempo down) and FFmpeg (compression) libraries.

\subsection{Training settings}
\label{sec:dataa}
We follow the same steps of data preparation and feature extraction as standard ECAPA-TDNN for the models we trained. The input features are the 80-dimensional MFCCs extracted from the window with 25ms length and 10ms shift, without voice activity detection. Three second random crops of the MFCCs feature vectors are normalized through cepstral mean subtraction. All models are trained using AAM-softmax \cite{10,11} with a margin of 0.2 and softmax prescaling of 30 for 4 cycles. We apply a weight decay on all weights in the model of 2e-4 incldue the AAM-softmax weights. We adopt 512 channels in the convolution frame layer in all the trained models. All models we trained using a cyclical learning rate varying between 1e-8 and 1e-3 using the triangular2 policy \cite{20} with the Adam optimizer \cite{21}. The duration of one cycle is set to 130k iterations. In the MACCIF-TDNN system for the encoder layer we employ 8 parallel attention layers. For each of these we use ${d_k} = {d_v} = 512/8 = 64$. For the multi-head attentive pooling we employ 2 parallel attention layers, and the parameter$\lambda$ and $\rho$ are set to 1 and 1 respectively. The mini-batch size for training is 64.
\subsection{Evaluation settings}
\label{sec:datac}
The scores are produced using the cosine distance between speaker embeddings extracted from the final fully connected layer for all systems. VoxCeleb1 dataset is used to evaluate the systems. The standard Equal Error Rate (EER) and minimum normalized detection cost function (MinDCF) is used as the evaluation metrics to the performance. For MinDCF calculation we assume ${P_{target}} = {10^{ - 2}}$ and ${C_{FA}} = {C_{Miss}} = 1$.

\subsection{Experimental Results}
\label{sec:ecapa}
To reveal how each of the proposed improvements affects the performance, we conduct a group of ablation studies and all the results are shown in Table 1. All ablation studies are trained with the training set of VoxCeleb1, and the testing set of VoxCeleb1 is used for evaluation. The experiment A.0 uses SE-Res2Block and Attentive statistic pooling(ASP) which is essentially the same to ECAPA-TDNN, we refer to the code public available in \cite {22} to complete this experiment and all the other improvements are based on this realization. The experiment A.1 uses multi-head ASP instead of the version applied in the experiment A.0. The experiment A.2 does not use the SE-Res2Block but the encoder layer of transformer. Different from the experiments A.1 and A.2, A.3 uses the integration structure of SE-Res2Block and transformer encoder module with ASP. Experiment A.4 uses all the proposed improvements which means the integration of SE-Res2Block,transformer encoder module and multi-head ASP. 
\begin{table}[h]\normalsize
\centering
\caption{Ablation study of the MACCIF-TDNN architecture}
\label{table2}
\begin{tabular}{lcc}
\hline 
\multicolumn{1}{c}{Architecture}     & EER(\%) & MinDCF \\ \hline
A.0:SE-Res2Block+ASP(ECAPA)     & 3.77    & 0.383  \\ \hline
A.1:SE-Res2Block+multi-head ASP    & 3.66    & 0.367  \\ \hline
A.2:encoder+multi-head ASP     & 3.73    & 0.371  \\ \hline
A.3:encoder+SE-Res2Block+ASP     & 3.62    & 0.365	  \\ \hline
A.4:MACCIF-TDNN     & \textbf{3.60}    & \textbf{0.362}  \\ \hline
\end{tabular}
\end{table} 

The comparision of the result of A.4 with A.2 confirms the benefits of the integration of SE-Res2Block which can improve the EER and MinDCF metric with 3.4\% and 2.4\%, respectively. This indicates that it is contributive to capture the channel interdependence and multi-scale local context features. The results of experiment A.4 and A.1 show that the encoder module of Transformer can lead to performance improvement about 1.6\% in EER and 1.3\% in MinDCF, respectively. This indicates that the limited local context of the frame-level features is insufficient and can be complemented with global utterance-based information. Comparing experiment A.4 with A.3, we can see that the multi-head attentive statistics pooling can make a improvement of performance about 0.5\% in EER and 0.8\% in MinDCF, respectively. This indicates that extending the pooling method in a multi-head way is helpful to discriminate features from multiple aspect. The above comparisions illustrate all the proposed improvements can bring positive impact on the performance of TDNN-based system. Additionally, the results of experiment A.4 and A.0 show that the aggregation of all improvements can achieve 4.5\% in EER and 5.4\% in MinDCF improvement, respectively. The overall improvement brings greater benefits than each single improvement reval that MACCIF-TDNN can indeed capture the complementary multi aspect channel and context features.

A performance overview of the baseline systems and our proposed MACCIF-TDNN system is given in Table 2. The ECAPA-TDNN(reproduced) and our proposed MACCIF-TDNN models are trained with the training set of voxceleb-2.  
Our proposed architecture MACCIF-TDNN can apparently the E-TDNN, R-vector and S-vector baseline systems, and gives a relative improvement of 4.8\% in EER and 3.2\% in MinDCF over the reproduced standard ECAPA-TDNN baseline system. 
This reveals that it is benefitial to aggregate the channel and context interdependence features multi aspect way in TDNN-based architecture can indeed bring performance improvement.
Despite, We do not reproduce the performance of standard ECAPA-TDNN(512) reported in \cite{7} using the public code available in \cite{22}. And either the ECAPA-TDNN(1024) or the latest ECAPA-CNN-TDNN is trained with 1024 channels in convolution frame layer, which can not be compared with MACCIF-TDNN directly. But, the MACCIF-TDNN can still achieve competitive results.

\begin{table*}[h]\normalsize
\centering
\caption{ EER and MinDCF performance of all systems on the standard VoxCeleb1 }
\label{table2}
\renewcommand{\arraystretch}{1.5}
\begin{tabular}{ccccccc}
\hline
\rule{0pt}{8pt}\multirow{2}{*}{\textbf{Architecture}} & \multicolumn{2}{c}{\textbf{VoxCeleb1}}       & \multicolumn{2}{c}{\textbf{VoxCeleb-E}} & \multicolumn{2}{c}{\textbf{VoxCeleb1-H}} \\ \cline{2-7} 
\rule{0pt}{8pt}                              & \textbf{EER(\%)} & \textbf{MinDCF}                    & \textbf {EER(\%)}        & \textbf{MinDCF}        & \textbf{EER(\%)}         & \textbf{MinDCF}        \\ \hline
\rule{0pt}{8pt}Standard ECAPA-TDNN(Reproduced)               & 1.25    & 0.153                     & 1.55           & 0.163         & 2.51            & 0.236         \\ \hline
\rule{0pt}{8pt}Standard ECAPA-TDNN               & 1.01    & 0.127                     & 1.24           & 0.142         & 2.32            & 0.218         \\ \hline
\rule{0pt}{8pt}ECAPA-TDNN(1024)              & 0.87                     & 0.107                     & 1.12  & 0.132 & 2.12  & 0.210 \\ \hline
\rule{0pt}{8pt}ECAPA CNN-TDNN                & 0.72 & 0.047                     & ----- & ----- & ----- & ----- \\ \hline
\rule{0pt}{8pt}E-TDNN \cite{1}                       & 1.49    & 0.160                     & 1.61           & 0.171         & 2.69            & 0.242         \\ \hline
\rule{0pt}{8pt}R-vector\cite{12}                      & 1.47    & 0.177                     & 1.60           & 0.179         & 2.88            & 0.267         \\ \hline
\rule{0pt}{8pt}S-vector\cite{3}                      & 2.67                     & 0.300 & ----- & ----- & ----- & ----- \\ \hline
\rule{0pt}{8pt}MACCIF-TDNN(512)              & \textbf{1.19}    &\textbf{0.148} &\textbf{1.47}           &\textbf{0.158}         &\textbf{2.48}            &\textbf{0.235}         \\ \hline
\end{tabular}
\end{table*}

\section{Conclusion}
\label{sec:ecapa}

In this paper, we propose a new network architecture which aggregates the channel and context interdependence features in a multi aspect way based on TDNN. Firstly, we use the SE-Res2Net blocks to explicitly model the channel interdependence, and process context features in a multi-scale way . Secondly, we explore to use the encoder structure of Transformer to model the long term context interdependence features. Before the pooling layer, we aggregate the outputs of SE-Res2Net blocks and Transformer-encoder to leverage the complementary channel and context interdependence features learned by themself respectively. Finally,  we extend the single attentive statistics pooling in a multi-head way which can discriminate features from multiple aspect. Experiments show that the proposed MACCIF-TDNN architecture can outperform most of the state-of-the-art TDNN-based systems on VoxCeleb1 test sets.

\bibliographystyle{IEEEbib}
\bibliography{refs}

\begin{thebibliography}{10}

\bibitem{1}
R.~Li, D.~Chen, and W.~Zhang, ``Voiceai systems to nist sre19 evaluation:
  Robust speaker recognition on conversational telephone speech,'' in {\em
  ICASSP 2020 - 2020 IEEE International Conference on Acoustics, Speech and
  Signal Processing (ICASSP)}, 2020.

\bibitem{2}
S.~Novoselov, A.~Gusev, A.~Ivanov, T.~Pekhovsky, A.~Shulipa, G.~Lavrentyeva,
  V.~Volokhov, and A.~Kozlov, ``Stc speaker recognition systems for the voices
  from a distance challenge,'' 2019.

\bibitem{3}
S.~Metilda, S.~V. Katta, and S.~Umesh, ``S-vectors: Speaker embeddings based on
  transformer's encoder for text-independent speaker verification,''

\bibitem{4}
A.~Vaswani, N.~Shazeer, N.~Parmar, J.~Uszkoreit, L.~Jones, A.~N. Gomez,
  L.~Kaiser, and I.~Polosukhin, ``Attention is all you need,'' {\em arXiv
  preprint arXiv:1706.03762}, 2017.

\bibitem{5}
D.~Snyder, D.~Garcia-Romero, D.~Povey, and S.~Khudanpur, ``Deep neural network
  embeddings for text-independent speaker verification,'' in {\em Interspeech
  2017}, 2017.

\bibitem{6}
D.~Snyder, D.~Garcia-Romero, G.~Sell, D.~Povey, and S.~Khudanpur, ``X-vectors:
  Robust dnn embeddings for speaker recognition,'' in {\em ICASSP 2018 - 2018
  IEEE International Conference on Acoustics, Speech and Signal Processing
  (ICASSP)}, 2018.

\bibitem{7}
B.~Desplanques, J.~Thienpondt, and K.~Demuynck, ``Ecapa-tdnn: Emphasized
  channel attention, propagation and aggregation in tdnn based speaker
  verification,'' in {\em Interspeech 2020}, 2020.

\bibitem{8}
J.~Thienpondt, B.~Desplanques, and K.~Demuynck, ``Integrating frequency
  translational invariance in tdnns and frequency positional information in 2d
  resnets to enhance speaker verification,''

\bibitem{9}
G.~Huang, Z.~Liu, V.~Laurens, and K.~Q. Weinberger, ``Densely connected
  convolutional networks,'' in {\em IEEE Computer Society}, 2016.

\bibitem{10}
J.~Deng, J.~Guo, and S.~Zafeiriou, ``Arcface: Additive angular margin loss for
  deep face recognition,'' in {\em 2019 IEEE/CVF Conference on Computer Vision
  and Pattern Recognition (CVPR)}, 2019.

\bibitem{11}
X.~Xiang, S.~Wang, H.~Huang, Y.~Qian, and K.~Yu, ``Margin matters: Towards more
  discriminative deep neural network embeddings for speaker recognition,'' in
  {\em 2019 Asia-Pacific Signal and Information Processing Association Annual
  Summit and Conference (APSIPA ASC)}, 2019.

\bibitem{12}
H.~Zeinali, S.~Wang, A.~Silnova, and O.~Plchot, ``But system description to
  voxceleb speaker recognition challenge 2019,''

\bibitem{13}
K.~Okabe, T.~Koshinaka, and K.~Shinoda, ``Attentive statistics pooling for deep
  speaker embedding,'' in {\em Interspeech 2018}, 2018.

\bibitem{14}
M.~India, P.~Safari, and J.~Hernando, ``Self multi-head attention for speaker
  recognition,'' in {\em Interspeech 2019}, 2019.

\bibitem{15}
Y.~Wu, C.~Guo, H.~Gao, X.~Hou, and J.~Xu, ``Vector-based attentive pooling for
  text-independent speaker verification,'' 2020.

\bibitem{16}
A.~Nagrani, J.~S. Chung, and A.~Zisserman, ``Voxceleb: a large-scale speaker
  identification dataset,'' in {\em Interspeech}, 2017.

\bibitem{17}
J.~S. Chung, A.~Nagrani, and A.~Zisserman, ``Voxceleb2: Deep speaker
  recognition,''

\bibitem{18}
T.~Ko, V.~Peddinti, D.~Povey, M.~L. Seltzer, and S.~Khudanpur, ``A study on
  data augmentation of reverberant speech for robust speech recognition,'' in
  {\em IEEE International Conference on Acoustics}, 2017.

\bibitem{19}
D.~Snyder, G.~Chen, and D.~Povey, ``Musan: A music, speech, and noise corpus,''
  {\em Computer Science}, 2015.

\bibitem{20}
L.~Smith, ``Cyclical learning rates for training neural networks,'' in {\em
  2017 IEEE Winter Conference on Applications of Computer Vision (WACV)}, 2017.

\bibitem{21}
D.~Kingma and J.~Ba, ``Adam: A method for stochastic optimization,'' {\em
  Computer Science}, 2014.

\bibitem{22}
``https://github.com/joovvhan/ecapa-tdnn,''

\end{thebibliography}

\end{document}